# Societal Adaptation to AI Human-Labor Automation

Yuval Rymon, Tel-Aviv University[1]

yuvalrymon@mail.tau.ac.il

November 2024

Abstract

*AI is transforming human labor at an unprecedented pace - improving 10x per year in training effectiveness. This paper analyzes how society can adapt to AI-driven human-labor automation (HLA), using Bernardi et al.'s societal adaptation framework. Drawing on literature from general automation economics and recent AI developments, the paper develops a "threat model". The threat model is centered on mass unemployment and its socioeconomic consequences, and assumes a non-binary scenario between full AGI takeover and swift job creation. The analysis explores both "capability-modifying interventions" (CMIs) that shape how AI develops, and "adaptation interventions" (ADIs) that help society adjust. Key interventions analyzed include steering AI development toward human-complementing capabilities, implementing human-in-the-loop requirements, taxation of automation, comprehensive reorientation of education, and both material and social substitutes for work. While CMIs can slow the transition in the short-term, significant automation is inevitable. Long-term adaptation requires ADIs – from education reform to providing substitutes for both the income and psychological benefits of work. Success depends on upfront preparation through mechanisms like "if-then commitments", and crafting flexible and accurate regulation that avoids misspecification. This structured analysis of HLA interventions and their potential effects and challenges aims to guide holistic AI governance strategies for the AI economy.*

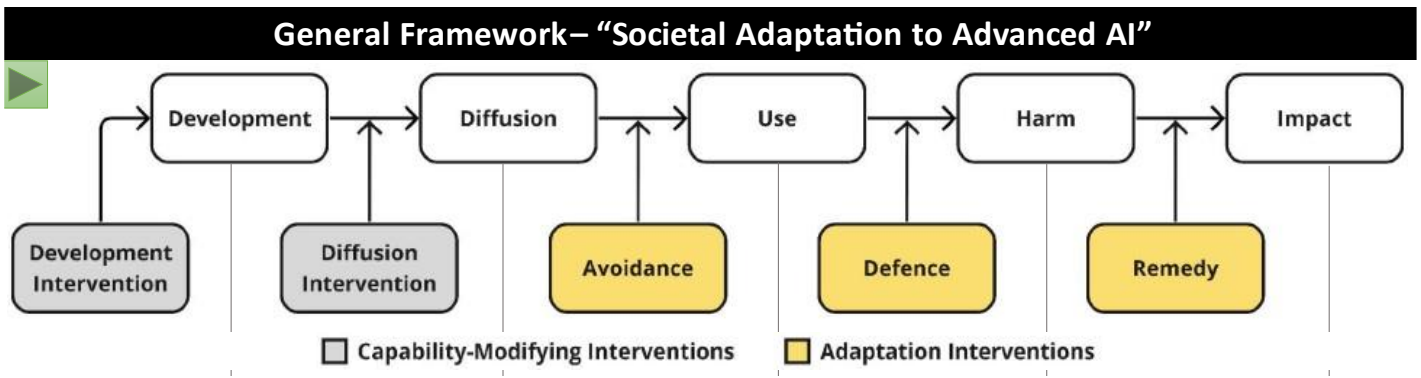

Source: "Societal Adaptation to Advanced AI", Bernardi et.al 2024

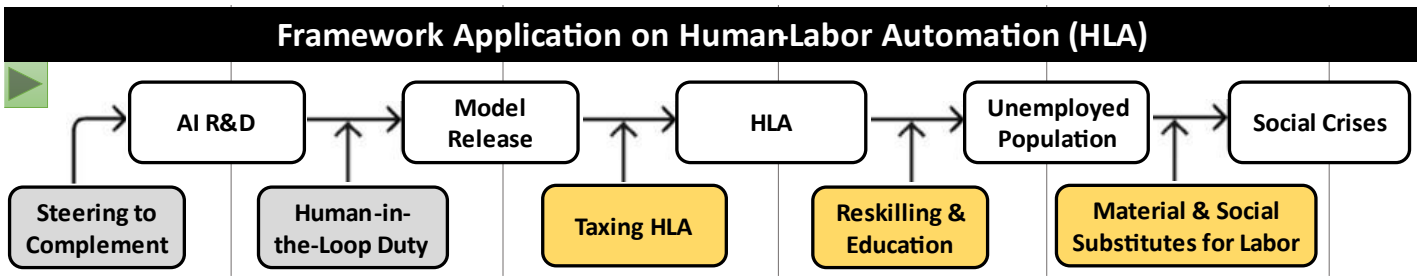

---

[1] An M.A. student in Political Science with a focus on Technology and Public Policy

# 1. Introduction

The threat of AI automating human labor isn't new, but today's AI is different. Unlike previous automation waves, modern AI can automate human tasks in an unprecedented scale and scope. Most of the popular discussions assume one of two extremes: either AGI makes all human labor obsolete, or the economy swiftly creates new jobs just like past technological revolutions. Reality in the coming decades likely lies between these poles – significant but partial automation that requires careful management.

This paper applies Bernardi et al.'s societal adaptation framework to analyze responses. The framework examines interventions across five stages (development, diffusion, use, initial harm, and impact) and distinguishes between capability-modifying interventions (CMIs) that shape how AI develops and adaptation interventions (ADIs) that help society adjust.

The paper proceeds as follows. Section 2 introduces Bernardi et al.'s societal adaptation framework, outlining its distinction between capability-modifying and adaptation interventions across five sequential stages. Section 3 examines general human-labor automation theory, analyzing effects within traditional economic models, transition dynamics and non-economic dimensions. Section 4 explores the distinct characteristics of AI-driven automation, its transformative potential, and current trajectories. Section 5 applies the societal adaptation framework to AI-driven HLA, developing a threat model and analyzing potential interventions across all framework stages. The paper concludes with main insights about the nature of the challenge, intervention effectiveness, implementation requirements.

# 2. The Societal Adaptation Framework

Bernardi et al.'s framework distinguishes between two categories of interventions addressing AI risk: capability-modifying interventions (CMIs) and adaptation interventions (ADIs). CMIs aim to identify harmful AI capabilities and modify their development and availability, while ADIs adjust societal structures to reduce the negative impacts of potentially harmful capabilities once created (Bernardi et.al, 2024, 2-3).

While the authors argue that CMIs become less effective over time[2], mainly due to increasing technological diffusion, they emphasize that CMIs remain a "crucially important component" of current regulation. Rather than replacing CMIs, they advocate for increased investment in complementary ADIs (Bernardi et.al, 2024, 2-3).

The framework's primary contribution is its structured analysis of the causal and chronological pathway through which AI capabilities generate negative societal impacts, and categorizing interventions at different points.

Implementation begins with constructing a "threat model" that identifies three elements: the potential harmful "use" of a technological capability, the immediate "harm" this use would generate, and the downstream "impact" on society. For instance, in analyzing cybersecurity risks, the "use" involves AI systems enabling small actors to conduct attacks, the "harm" comprises infrastructure compromise or data theft, and the "impact" encompasses economic damage, national security implications, and potential loss of life (Bernardi et.al, 2024, 6).

The framework delineates five sequential stages with corresponding intervention categories (Bernardi et.al, 2024, 3-5):

---

[2] This is due to increased diffusion of ever more cheap and efficient models to masses of actors, which undermines the feasibility of governing capabilities. Additionally, they argue that AI safeguards are not fail-safe, meaning they can easily be circumvented in a variety of ways that would not be amenable. Even if CMI could continue to work well, they point out that they stop innovation by preventing beneficial uses of potentially harmful capabilities, as opposed to ADI which focus on applications alone.

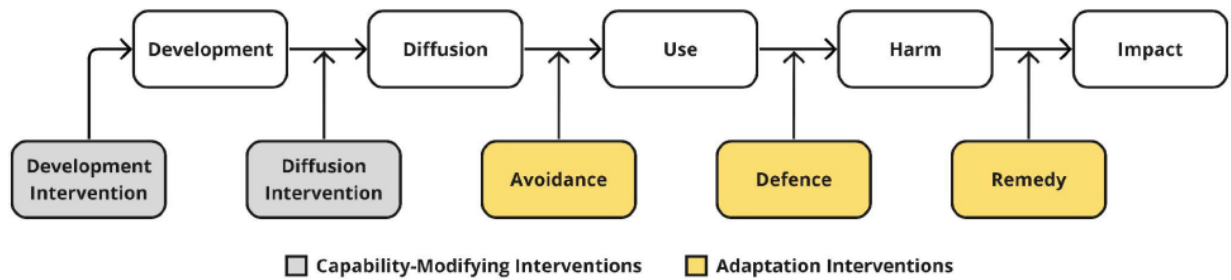

*Source: "Societal Adaptation to Advanced AI", Bernardi et.al 2024*

⇒ **Development interventions:** CMIs affecting which capabilities are developed, preceding the development stage. For example, companies could refrain altogether from developing systems that have potentially harmful capabilities, or make them more resistant to hacking.

1. **Development Stage:** an AI capability or system is developed.

⇒ **Diffusion interventions:** CMIs affecting which capabilities are made available, to whom and with what degree of access. For example, companies can employ "staged release", allow access only via an API with secure safeguards, or enforce terms of service policies.

2. **Diffusion Stage:** the capability or system becomes available to a wide crowd of users.

⇒ **Avoidance interventions:** ADIs reducing the expected extent of the potentially harmful use, by making such use more difficult to engage in, or more costly. For example, this can be done by limiting the user's access to key resources/actors that are required for the activity, or by building institutions with credible punishment threats.

3. **Use Stage:** the AI system is used in a way that could cause harm. This harm could be actively intended ("misuse"), or unintentional ("accident").

⇒ **Defense interventions:** ADIs reducing the extent of initial harm, after the potentially harmful use occurred. In spearphishing, "defence" is a matter of reducing the chance that malicious emails succeed in giving the cybercriminal access to sensitive information (for example by educating the users, or hardening auto protections on their computers).

4. **Initial Harm Stage:** The harmful use of AI results in an harmful outcome (like gaining access to sensitive information).

⇒ **Remedial interventions:** after the initial harm occurs, remedial ADIs can reduce or eliminate the expected negative impact downstream. In the spearphishing example, this can be done by reducing the extent to which company secrets are undermined in case of information stolen by an external actor. For instance, companies can include some false and misleading documents on its servers.

5. **Impact Stage:** The initial harm results in further negative impact downstream. This impact could be measured in terms of e.g. lives lost, economic opportunities lost, or damage to national security.

# 3. General Human-Labor Automation

The dominant theoretical model in human-labor automation (HLA) literature is the task-based model, which conceptualizes economic production as a process of completing various tasks performed by humans, machines, or both. Within this model, automation is defined as the introduction of machinery to perform tasks previously executed by human labor (Acemoglu and Restrepo, 2019, 198-200). This process affects labor through multiple distinct channels.

## 3.1 Displacement and Countervailing Effects

The primary impact is the **Displacement Effect** – a reduction in labor demand, wages, and employment resulting from the diminished exclusive domain of human capabilities (Acemoglu and Restrepo, 2019, 198). However, Acemoglu and Restrepo identify several countervailing effects, which can balance at least part of the displacement effect:

- **Productivity Effect:** The automation of tasks reduces production costs, potentially decreasing prices for goods and services. This increase in household disposable income[3] can stimulate demand for some goods and services, subsequently raising labor demand in either or both automated and non-automated industries – depending on areas of increased demand for consumption. However, this effect's efficacy diminishes if automation's economic benefits concentrate among a small segment of the population, such as when market power enables firms to maintain prices despite reduced costs (Acemoglu and Restrepo, 2019, 198, 203, 228).
- **Deepening of Existing Automation:** Technological improvements in already-automated tasks enhance capital productivity without displacing additional labor, as these tasks no longer employ human workers. For example, if a computer can now perform the same calculations but more efficiently, no employee is currently in charge of calculating manually and no one can be fired. This productivity enhancement reinforces the Productivity Effect without generating additional displacement (Acemoglu and Restrepo, 2019, 204).
- **Capital Accumulation** As automation increases capital's role in production, the resulting higher demand for capital elevates rental rates (interest on loans for example), leading to capital accumulation. This accumulated capital creates new investment opportunities, generating increased labor demand for new endeavors (Acemoglu and Restrepo, 2019, 204).
- **Reinstatement Effect:** The most significant countervailing force is the creation of new tasks where human labor maintains comparative advantage. This occurs through two mechanisms:
    1. The Displacement Effect creates a pool of less expensive labor, making previously unprofitable labor-intensive tasks economically viable.
    2. New automation technologies serve as platforms for creating novel human tasks.

    Historical periods of intensive automation, such as the industrial and agricultural revolutions, demonstrate this pattern of job creation alongside technological advancement (Acemoglu and Restrepo, 2019, 198-199, 205-207).

## 3.2 Transition Dynamics and Social Implications

The societal impact of automation extends beyond the interplay of these economic effects to encompass also the transition's speed and nature. Worker reallocation to new tasks historically proves slow and painful. It takes time for workers to find new jobs and tasks in which they can be productive, and interim unemployment can create depressed labor markets that compound adjustment costs (Acemoglu and Restrepo, 2019, 208-209).

The British Industrial Revolution exemplifies these dynamics: despite eventually generating increased labor demand and wages, it precipitated 80 years of wage stagnation, expanding poverty, and harsh living conditions. The transition's resolution coincided with the implementation of mass education, which enhanced workforce capabilities and facilitated new task creation (Acemoglu and Restrepo, 2019, 208-209).

---

[3] They can now spend the money they saved due to the lowering of original prices.

The lesson Acemoglu and Restrepo highlight is that the creation of new tasks is not an autonomous process. It is shaped by decisions on issues such as labor skilling, but also by investment in technologies that complement labor, and tax policies influencing the labor-machine cost-benefit. The political economy aspect is key as the public has the power to slow down or even stop technological developments, if prosperity is not shared across wide segments of society (Acemoglu and Restrepo, 2019, 228).

### 3.3 Non-Economic Dimensions

While discourse on automation's impact often centers on wage loss, Dijekma and Gunderson's research synthesis reveals critical non-monetary benefits of employment:

- **Socialization** – Community membership, interpersonal relationships, and navigating organizational structures.
- **Identity** – Both external recognition and internal self-conception (Dijekma and Gunderson, 2019, 5-6).
- **Meaning and self-esteem** – Sense of purpose and self-worth with contribution to satisfaction (Dijekma and Gunderson, 2019, 5-6).
- **Cognitive development** – Through workplace challenges and problem-solving (Dijekma and Gunderson, 2019, 12).
- **Health** – the link between physical health and mental health and unemployment is strong. The reasons associated with this are the social negative perception of unemployment and the individual loss of benefits and healthy lifestyle routines (Dijekma and Gunderson, 2019, 8-9).

These non-monetary benefits are so substantial that some scholars argue *"the main cost of job loss is psychological"* (Dijekma and Gunderson, 2019, 15). Even if potentially inflated, this assertion underscores the necessity of incorporating psychological impacts into HLA threat models. It suggests individuals may seek employment even with significantly reduced wages, highlighting work's intrinsic value beyond financial compensation.

## 4. AI-Driven Human-Labor Automation

While historical patterns of human-labor automation (HLA) provide valuable insights, artificial intelligence presents fundamentally distinct characteristics that warrant separate analysis.

### 4.1 Defining Contemporary AI

Contemporary definitions from major institutions emphasize AI's broad scope and capabilities. The OECD characterizes AI as a *"machine-based system that for explicit or implicit objectives, infers, from the input it receives, how to generate outputs such as predictions, recommendations, content, or decisions that can influence physical or virtual environments. Different AI systems vary in their levels of autonomy and adaptiveness after deploymen."*. Complementarily, the EU AI Act and US Executive Order on Safe, Secure, and Trustworthy AI define it as a *"loose term used to describe a range of advanced technologies that exhibit human-like intelligence including machine learning, autonomous robotics and vehicles, computer vision, language processing, virtual agents, and neural networks"* (Filippucci et al., 2024, 8).

Advanced AI systems, such as Large Language Models (LLMs) like ChatGPT, are capable of performing unprecedently broad and advanced tasks previously exclusive to humans. Their autonomous operation and capacity for self-improvement, coupled with rapid development trajectories, suggest profound implications for labor markets (Filippucci et al., 2024, 23).

## 4.2 Transformative Potential

There is a lot of uncertainty about the pace of change and the magnitude of the coming disruption. Gruetzemacher and Whittlestone delineate three key elements of transformative societal change from a technology: **Practical Irreversibility**: Technology becomes so deeply embedded that society cannot feasibly reverse its adoption and use; **Breadth**: Impact extends across multiple domains of life rather than remaining confined to specific sectors; and **Extremity**: Changes fundamentally alter metrics of human progress and well-being. These elements guide a taxonomy of transformative technologies (Gruetzemacher and Whittlestone, 2022, 7-10):

- **Narrow Transformative Technology** – Exhibits Practical Irreversibility, but lacks Breadth and Extremity. Example: nuclear weapons, which changed warfare completely but did not influence other domains.
- **General-Purpose Technology" (GPT)** – Demonstrates Practical Irreversibility and Breadth, driving change across many or all domains with applicability to a wide variety of tasks. It eventually leads to a significant increase in economic productivity. However, it lacks Extremity in isolation as no single invention in history changed the metrics of human progress alone. Example: electricity, which led to substantial changes in daily life and communication, enabling products we take for granted like light bulbs and telephones.
- **Revolutionary/Radical Transformative Technology** – Incorporates all three elements. Example: agricultural and industrial revolutions, both constituting clusters of technological innovations and producing unprecedented societal transformations. The Agricultural Revolution represents the transition from people living as hunter-gatherers to large, settled civilizations. The Industrial Revolution represents the transition to mechanized manufacturing and factories, leading to unprecedented population growth and rising quality of life, which particularly coincided with changes in metrics of human well-being (physical health, economic well-being, energy capture and technological empowerment).

AI is becoming a GPT, but it is still unclear whether it will end up precipitating fundamental and unprecedented societal change on the level of the industrial revolution. Its distinctiveness lies in functioning not as a singular technology but as a method generating multiple technological capabilities like natural language processing, computer vision, and robotic learning (Gruetzemacher and Whittlestone, 2022, 13-15). Advanced AI capable of performing physical and intellectual tasks at or above human level – often termed Artificial General Intelligence (AGI) – could fundamentally alter the dynamics of our economic system, and in the context of HLA create the potential for widespread labor displacement (Korinek, 2024, 6).

## 4.3 Current Trajectory and Limitations

Recent AI development during the last decade exhibits extraordinary growth metrics. AI's growth was pushed by an exponential increase in the effectiveness of training AI models[4] – 10 times per year (!) over the past decade. This is thanks to a 4-times per year increase in the amount of computer power employed (powered by Moore's law and a tripling of investments each year), and a 2.5-times per year increase in the algorithmic efficiency of utilizing computing power (Korinek, 2024, 4-5).

The AI community has developed "scaling laws" which are the predicted regularities (holding for the past decade and used by AI companies) that describe of how much better AI becomes in prediction as the number of model parameters (representing complexity) and training data increase (Korinek, 2024, 5-6). If current trajectories continue in the future, AI is likely to dominate many more human tasks.

---
[4] An optimization process of prediction by receiving more data as input and feedback on outputs.

However, several factors suggest caution in extrapolating these trends. First, these "scaling laws" are not actual laws, but very short-term regularities with no strong evidence to support the claim they will continue to hold. Second, the current architecture of Artificial Neural Networks[5], aiming to emulate the human brain's neural system, is extremely simplistic and lacking. Worse than that, our current understanding of how the brain processes information is extremely limited. This is partly due to the lack of tools to record, stimulate, measure and model all we wish – which constrain concepts and theories[6] (Carlsmith, 2020). In light of this, it is unreasonable to assume that an extreme simplification (as artificial neural networks are) captures everything that matters in the brain's biophysical complexity, and conjecture we can get 100% of humans' cognitive abilities with just scaling computer power on current architecture.

In the short-term, it seems likely that AI will facilitate the creation on new tasks. Such new tasks will (and in fact already do) include the training, monitoring, explaining and sustaining of models. But additionally, Acemoglu and Restrepo suggest that the personalized customization AI allows to (currently) uniform and universal services like healthcare and education can create many more jobs. These jobs could consist of monitoring, designing and implementing individualized programs (Acemoglu and Restrepo, 2019, 207).

Even in the medium-term, assuming AI does reach a human-level in most or all tasks, there are factors that will sustain the demand for human labor in some areas. Korinek divides them between temporary social barriers (1-3), and fundamental human-centric aspects (4-7), and they are summarized below (Korinek, 2024, 13-15):

1. **Production and diffusion lags:** even when AI systems cost and perform as humans do, adoption could be postponed due to social and relationships between humans (employers and employees, consumers and producers), until inefficiency is too large, or competitive forces are strong enough.
2. **Generational Trust Differences:** older generations may be less comfortable with AI services.
3. **Regulatory Protection of Certain Professions:** currently protected jobs include politicians, lawmakers, judges, doctors, etc. They may remain only as a human facade for AI decision-making.
4. **Demand for Authentic of human interaction:** even if the machines are objectively indistinguishable or superior.
5. **Human identification:** competitive or performative jobs in sports and the arts, are centered around the idea that spectators can identify with a fellow human that pushes human limits.
6. **Religious Requirements** current religious beliefs require some acts to be performed by humans since adherents believe that humans distinguish themselves from all other entities.
7. **AI alignment:** Even in a world of AGI, AI alignment[7] will require a human perspective since only humans can decide what constitutes "aligned" behavior, based on their preferences and lived experience (and not intelligence).

---

[5] An Artificial Neural Network (ANN) is a system made up of layers of connected nodes, like a web. The first layer takes in the data, the middle layers process it, and the last layer gives the result. Each connection has numbers (weights and biases) that control how strongly data flows. The network learns by comparing its results to the correct answer, adjusting the connections step by step to improve. This process is repeated many times until the network gets better.

[6] Joseph Carlsmith shares a few quotes from neuroscientists that represent this:
"Despite decades of intense research efforts…we do not have any well-grounded, and certainly not generally accepted, theory about how neurons work together to provide the salient brain functions." – Einevoll et al., 2015.
"Neuroscience is collection of facts, rather than ideas; what is missing is connective tissue… We think we know roughly what neurons do…but not what they are communicating…we know almost nothing about what those motifs [repeated structures at the neocortex – YR] are for, or how they…support complex real-world behavior…we are still at a loss to explain how the brain does all but the most elementary things. We simply don't understand how the pieces fit together" – Gary Marcus, 2015.

7 Ensuring AI systems behave in accordance with human values and intentions.

# 5. Human-Labor Automation Societal Adaptation Framework

Having established the theoretical foundations of AI-driven HLA, we now apply Bernardi et al.'s societal adaptation framework to analyze potential interventions.

## 5.1 Threat Model

- **Use:** The continuous advancement in computational power and algorithmic efficiency suggests AI systems may automate a substantial portion of human cognitive and physical tasks across sectors. This analysis assumes a "non-binary" scenario – neither complete AGI dominancer endering human labor obsolete nor swift task creation – where AI's impact on human labor is significant enough to necessitate a substantial societal response.
- **Harm:** The automation of significant portions of the current labor market, coupled with displacement effects exceeding countervailing effects[8], may precipitate sharp decreases in employment and wages and loss of work's non-monetary benefits to individual and societal well-being[9].
- **Impact:** mass HLA could generate extreme and growing economic inequality between few automation beneficiaries (technology companies, capital owners, highly skilled workers) and the displaced workforce left unemployed and poor. Such extreme inequality together with poverty and joblessness could to lead to political unrest, weakened social cohesion and challenges to democratic governance like populist movements or elite-authoritarianism[10].

## 5.2 Development Interventions

**Steering to Complement:** a CMI designed to designed to promote human-complementing AI development that can increase labor demand while discouraging human-replacement technologies. Stiglitz and Korinek propose concrete measures such as reducing labor taxation, Subsidizing labor-friendly research through government funding criteria and IP laws, and incorporating union input in R&D investment decisions (Korinek and Stiglitz, 2020).

**While this intervention may temporarily slow automation, market forces will likely prevail. The potential productivity gains from autonomous systems create overwhelming economic incentives that will eventually overcome public resistance and counter-incentives.**

## 5.3 Diffusion Interventions

**Human-in-the-loop Duty:** a CMI requiring human oversight in algorithmic decision processes, initially in high-stakes domains (politics, law, healthcare, transportation) with potential expansion to other sectors (at least as conventional standards) comprehensively run by AI systems. Crootof et.al. identify several HITL roles (Crootof et.al., 2023, 473-487):

- Corrective: error, bias and situational correction.
- Resilience: backup and emergency override.
- Justificatory: explainability and transparency.
- Dignitary: preserving human dignity and enabling participation in decisions.
- Accountability: taking legal liability.
- Stand-in: purely symbolic with no influence.
- Friction: slowing down automated processes with human bottlenecks.
- Warm-body: protecting human jobs over performance.

---

[8] The productivity effect, the deepening of existing automation, capital accumulation, and the creation of new tasks.

[9] Such as socialization, identity, meaning and self-esteem, cognitive development and physical and mental health.

[10] Additionally, as nations with differing AI capabilities or automation policies gain or lose dramatic economic and political power, it can also threaten the global order. In this article I aim to focus on nation-states, and therefore will not analyze solutions to this challenge.

- Interface: translating and mediating between system and users/customers.

**While this intervention may create new human-exclusive tasks, these roles likely cannot absorb the majority of displaced workers.**

### 5.4 Avoidance Interventions

**Taxing Human-Labor Automation:** an ADI increasing automation costs to incentivize labor demand and potentially raising fundings for labor-friendly policies. A key implementation challenge centers on misspecification of HLA. Bullock and Winter warn that AI governance is especially vulnerable to misspecification due to the complexity and opacity of systems which makes their abilities unpredictable and the rapid advancements that can make regulations quickly obsolete (Bullock and Winter, 2024, 19-20). They analyze case-studies representing two types of misspecifications (Bullock and Winter, 2024, 5-6):

1. Overinclusive – stifling unrelated and beneficial actions. Excessive constraints may impede beneficial innovation, and potentially even stop labor-friendly complementary technologies.
2. Underinclusive – allowing loopholes to bypass regulation. Rapid evolution of HLA technologies may enable regulatory circumvention to avoid taxation.

In light of this challenge, adopting an HLA tax should be carefully crafted. Bullock and Winter recommend minimizing the usage of proxy metrics as ends, stating objectives explicitly to ease interpretations and updating, allowing rapid updates and action, regularly reviewing effectiveness, and designing flexibly to allow evolution (Bullock and Winter, 2024, 15-18).

**While it can also act as an income source for redistribution to labor-friendly policies, its misspecification challenge is serious. This suggests that implementation may require an overinclusive approach, despite economic costs.**

### 5.5 Defense Interventions

**Reskilling and Education:** an ADI providing the workforce with suitable skills to perform remaining and newly created human-exclusive tasks. In addition, a skilled population could promote the process of new task creation by producing more innovation. Korinek argues for a "fundamental reorientation" of educational curricula if an AI dominating many human tasks emerges. He suggests this reorientation should revolve around AI literacy (capabilities and limitations) and use (integration, interpretations and decision-making) in the short-term, and on remaining human-centric aspects (authenticity, identity, and oversight) in the long-term (Korinek, 2024, 16-17).

**This policy can significantly ease and expedite the "painful adjustment" phase during the transition towards an AI human-labor market. Additionally, it may undermine current educational inequalities, if current skills and "quality signals" become obsolete. Promoting a "fundamental reorientation" is likely to pay off in the present, as AI is already becoming a GPT with current capabilities. To achieve a high level of AI literacy across society, a comprehensive reform reaching beyond traditional school-age education is required.**

### 5.6 Remedial Interventions

**Material Substitutes for Labor:** ADIs replacing traditional labor financial income.

- **Universal Basic Income (UBI):** paying every citizen with an unconditional basic income on a permanent basis, that can provide "basic needs" (an ambiguous definition). Current objections to UBI mainly emphasize its potential to reduce the incentive to work, and the large tax costs of implementing it. However, in an AI-driven HLA economy, very few will be able to work. Moreover, it is hoped among tech leaders like Altman that the material abundance created will allow supporting the current basic standard of living. His proposed tax sources to replace labor income tax are companies' shares and privately owned land (Altman, 2021). For such an action to be stable, it would need to be assured of long-term survival outside the influence of day-to-day politics, and pre-legislated "behind a veil of ignorance", before conflicts of interest intensify in real time. A potential mechanism is "if-then commitments", proposed by Karnofsky, according to which, if condition X is met, then measures Y are implemented (Karnofsky, 2024, 1). Specifically, one such proposal related to economic inequality is the "windfall clause". According to this proposal, *"firms would bindingly agree to donate a meaningful portion of their profits if they earn a historically unprecedented economic windfall from the development of advanced AI"*, where "windfall" is largely defined as an income worth of some 1% of global GDP (O'Keefe et al., 2020, 2).

- **Universal Basic Capital (UBC):** a similar proposal to UBI, UBC is defined by Le Grand as a one-time large grant to every citizen, instead of periodically small amounts like in UBI (Le Grand, 2020, 1). The main difference may be the independence it allows in choosing how to invest actively, compared of receiving a passive source of income. While this difference could incentivize better investments, it is also more prone to misuse and losses that make people unable to support themselves. A possible solution is combining UBI as a guaranteed floor and UBC as a chance to be "part of the game" in an AI economy.

**Social Substitutes for Labor:** ADIs maintaining work's non-monetary benefits[11].

- **Unpaid Social Work:** providing all of work's non-monetary benefits.
- **Community Activities:** providing socialization experiences.
- **Political Direct Participation**: providing meaning and cognitive engagement.

**Material interventions require sustained economic growth and political stability, necessitating early commitment mechanisms. Social substitutes represent potential cultural adaptations rather than formal policies, yet remain crucial for maintaining social cohesion and individual well-being.**

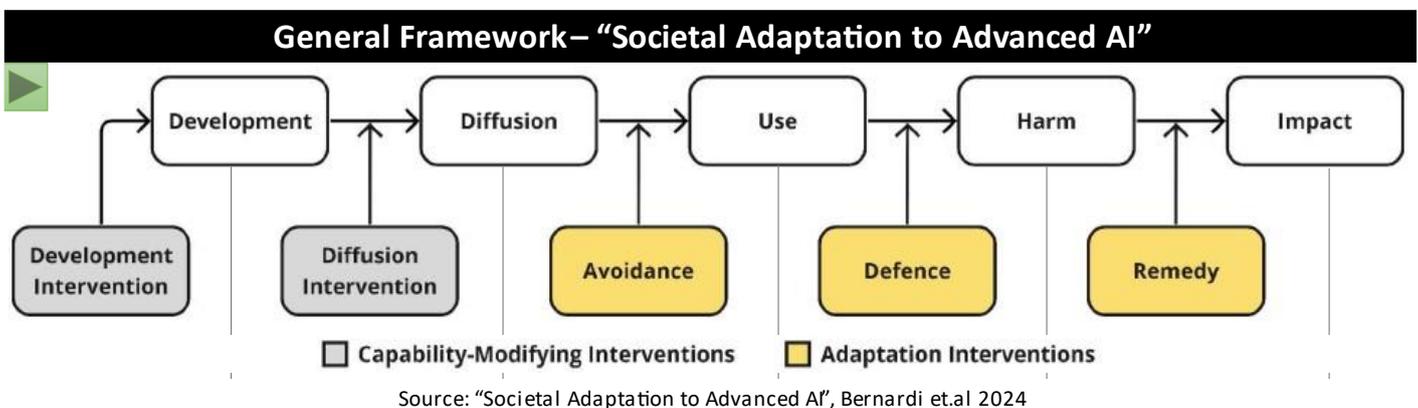

Source: "Societal Adaptation to Advanced AI", Bernardi et.al 2024

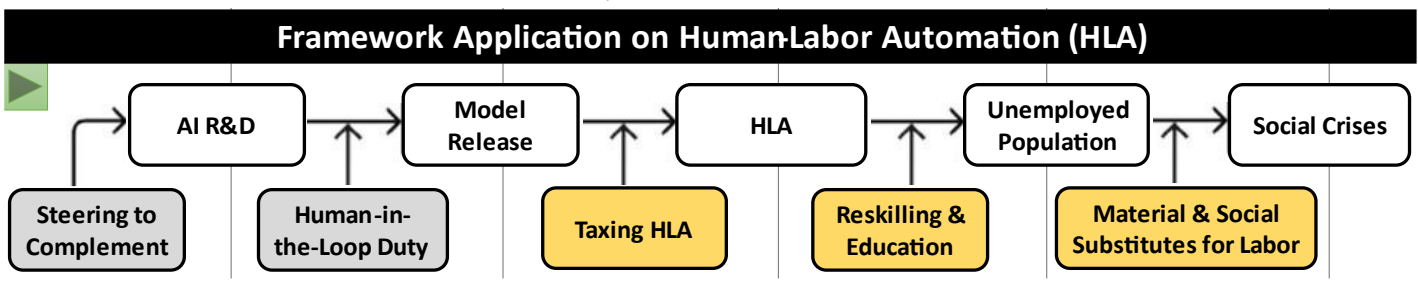

---

[11] Socialization, identity, meaning and self-esteem, cognitive development and health.

# 6. Conclusion

The challenge of AI-driven HLA requires a comprehensive strategy that both manages the transition and prepares for a fundamentally different economic and social paradigm. This paper's application of Bernardi et al.'s societal adaptation framework to AI-driven HLA provides a structured analysis of potential responses across different intervention stages. The analysis reveals several insights about the nature of the challenge, intervention effectiveness, implementation requirements.

## 5.1 Nature of HLA Change

- **Unprecedented Scale and Scope**: Unlike previous technological revolutions, AI-driven automation has an extraordinary pace of development (10x improvement per year in training effectiveness), and breadth of potentially automatable tasks (spanning both physical and cognitive domains). While it is useful to draw parallels with previous revolutions, these crucial differences necessitate new concepts and frameworks for understanding the current transition.
- **Non-Binary Reality**: The most probable scenarios appear to be those who lie between two "extremes": complete AGI takeover rendering all human labor obsolete, and swift and comprehensive creation of new jobs. These "middle ground" scenarios require careful consideration of partial automation effects and graduated response strategies.
- **Dual Impact of Job Loss**: Unemployment from HLA creates a material deprivation from loss of wages. However, a more neglected but perhaps equally important impact is the psychological and social effects from loss of work's intrinsic benefits (identity, meaning, socialization, etc.).

## 5.2 Intervention Effectiveness

- **Capability-Modifying Interventions (CMIs)**: such as steering development toward human-complementing AI and implementing human-in-the-loop requirements, have short-term effectiveness in slowing automation. Nevertheless, their viability is limited in the long-term due to competitive pressures and growing efficiency differentials.
- **Adaptation Interventions (ADIs)**: such as educational reorientation, and material (UBI/UBC) and social substitutes for work (taxation of HLA is technically an ADI but has similar effects to CMIs) allow a more robust societal adjustment.

## 5.3 Critical Success Factors

- **Early Preparation**: by pre-emptive measures such as "if-then commitments", that can establish triggers and metrics for intervention activation, while stakes are still moderate. Future work should examine these metrics and triggers.
- **Regulatory Design**: avoiding over-inclusive and under-inclusive misspecification, with adaptive frameworks that can evolve with technological change
- **Holistic Approach**: coordinating across intervention stages, balancing between slowing transition and preparing for it, and addressing both material and non-material work aspects.
- **Education Reform:** beginning a reorientation toward AI literacy and human-centric skills at present, as this offers value regardless of automation's exact pace.